\documentclass
[superscriptaddress,secnumarabic,assymb,amsmath,nobibnotes,aps,prd,showkeys,showpacs,nofootinbib,onecolumn,notitlepage,12pt]{revtex4}%
\usepackage{setspace}
\usepackage{color}
\usepackage{amsmath}
\usepackage{amsfonts}
\usepackage{verbatim}
\usepackage{amssymb}
\usepackage{graphicx,bm}
\usepackage{graphicx}
\usepackage{bbm}
\usepackage{amsmath}
\usepackage{amssymb}
\usepackage{bbm}
\usepackage{amssymb}
\usepackage{accents}
\usepackage{graphicx,bm}
\usepackage{graphicx}
\usepackage{bbm}
\usepackage{epstopdf}
\usepackage[caption=false]{subfig}
\usepackage{makecell}
\usepackage{hyperref}%
\providecommand{\U}[1]{\protect\rule{.1in}{.1in}}

\newcommand{\be}{\begin{equation}}
\newcommand{\ee}{\end{equation}}

\newcommand{\mincir}{\raise
-3.truept\hbox{\rlap{\hbox{$\sim$}}\raise4.truept\hbox{$<$}\ }}
\newcommand{\magcir}{\raise
-3.truept\hbox{\rlap{\hbox{$\sim$}}\raise4.truept\hbox{$>$}\ }}

\ifx\pdfoutput\relax\let\pdfoutput=\undefined\fi
\newcount\msipdfoutput
\ifx\pdfoutput\undefined\else
\ifcase\pdfoutput\else
\ifx\paperwidth\undefined\else
\ifdim\paperheight=0pt\relax\else\pdfpageheight\paperheight\fi
\ifdim\paperwidth=0pt\relax\else\pdfpagewidth\paperwidth\fi
\fi\fi\fi
\begin{document}
\title{Kantowski-Sachs and Bianchi III dynamics in $f\left(  Q\right) $-gravity}
\author{Alfredo D. Millano}
\email{alfredo.millano@ce.ucn.cl}
\affiliation{Departamento de Matem\'{a}ticas, Universidad Cat\'{o}lica del Norte, Avenida
Angamos 0610, Casilla 1280 Antofagasta, Chile}
\author{K. Dialektopoulos}
\email{kdialekt@gmail.com}
\affiliation{Department of Mathematics and Computer Science, Transilvania University of
Brasov, 500091, Brasov, Romania}
\author{N. Dimakis}
\email{nikolaos.dimakis@ufrontera.cl}
\affiliation{Departamento de Ciencias F\'{i}sicas, Universidad de la Frontera, Casilla
54-D, 4811186 Temuco, Chile}
\author{A. Giacomini}
\email{alexgiacomini@uach.cl}
\affiliation{Instituto de Ciencias F\'{\i}sicas y Matem\'{a}ticas, Universidad Austral de
Chile, Valdivia 5090000, Chile}
\author{H. Shababi}
\email{h.shababi@scu.edu.cn}
\affiliation{Center for Theoretical Physics, College of Physics, Sichuan University, Chengdu 610065, P. R. China}
\author{Amlan Halder}
\email{amlanhalder1@gmail.com}
\affiliation{School of Technology, Woxsen University, Hyderabad 502345, Telangana, India}
\author{A. Paliathanasis}
\email{anpaliat@phys.uoa.gr}
\affiliation{Institute of Systems Science, Durban University of Technology, PO Box 1334,
Durban 4000, South Africa}
\affiliation{Departamento de Matem\'{a}ticas, Universidad Cat\'{o}lica del Norte, Avenida
Angamos 0610, Casilla 1280 Antofagasta, Chile}
\affiliation{School of Technology, Woxsen University, Hyderabad 502345, Telangana, India}

\begin{abstract}

We explore the phase-space of homogeneous and anisotropic spacetimes within symmetric teleparallel $f(Q)$-gravity. Specifically, we consider the Kantowski-Sachs and locally rotational Bianchi III geometries to describe the physical space. By analyzing the phase-space, we reconstruct the cosmological history dictated by $f(Q)$-gravity and comment about the theory's viability. Our findings suggest that the free parameters of the connection must be constrained to eliminate nonlinear terms in the field equations. Consequently, new stationary points emerge, rendering the theory cosmologically viable. We identify the existence of anisotropic accelerated universes, which may correspond to the pre-inflationary epoch.

\end{abstract}
\maketitle
\date{day}

\section{Introduction}

Even though the flat Friedmann-Lema\^itre-Robertson-Walker (FLRW) $\Lambda$
Cold Dark Matter ($\Lambda$CDM) model passes most of the observational tests
with flying colours, recent cosmological data suggest a tension between early
and late Universe measurements of the expansion rate, i.e. the Hubble
parameter. Specifically, Cosmic Microwave Background (CMB) data from Planck
collaboration in their 2018 release \footnote{Currently available online in
https://wiki.cosmos.esa.int/planck-legacy-archive/index.php/Cosmological\_Parameters
(accessed on September 22nd, 2023).} \cite{Planck:2018vyg}, propagated to
today using flat $\Lambda$CDM suggest a value of the Hubble parameter at
$H_{0} = 67.4\pm0.5$ km/s/Mpc. On the other hand, Cepheid variable stars are
used as standard candles to measure the distances of galaxies with the
distance ladder method, inferring a value of $H_{0} = 74\pm1.4$ km/s/Mpc (see
SH0ES \cite{Riess:2019cxk}). The statistical significance between these two
measurements is at 5$\sigma$. In addition, a nontrivial tension is related to
the amplitude of density fluctuations at low redshifts when compared to the one predicted by CMB. The
amplitude of these fluctuations, also known as linear matter perturbations, is
often defined by the value of the linear matter overdensity field in spheres
with a radius $8 h^{-1}$Mpc. This value is called $\sigma_{8}$. The parameter
$S_{8}$ is defined as $\sigma_{8} \left( \Omega_{m}/0.3\right) ^{\alpha}$,
where $\Omega_{m}$ is the fractional energy density of nonrelativistic matter
and $\alpha$ is selected to minimize the correlation between $S_{8}$ and
$\Omega_{m}$. Several weak lensing surveys such as DES \cite{DES:2017qwj}, HSC
\cite{HSC:2018mrq}, and Heymans et al. \cite{Heymans:2020gsg}, have measured
$S_{8}$, but they have encountered different levels of disagreement with the
value inferred from measurements by the Planck satellite. Cosmic shear data tend to recover slightly lower values of $S_{8}$ than the CMB. The
Kilo-Degree Survey collaboration (KiDS) has reported the most significant
disagreement so far with a significance of around 3$\sigma$
\cite{Heymans:2020gsg, Hildebrandt:2020rno, KiDS:2020suj, KiDS:2020ghu,
Garcia-Garcia:2021unp}. These are only some of the issues that the concordance
cosmological model needs to address.

A possible alternative to solve these tensions is to consider extensions
beyond $\Lambda$CDM; see \cite{DiValentino:2021izs} for a review. However,
incomprehension between the SNIa absolute magnitude and the Cepheid-based
distance ladder instead of an exotic late- or early-time physics could be the
reason for the tensions \cite{Efstathiou:2021ocp}.

Because of the above, even though inflation is widely considered to be the
most plausible explanation for the homogeneity and isotropy of the observable
Universe, among others, most of the studies base their analysis on the fact
that the Universe should be flat FLRW and, based on that, they study the
evolution of perturbations. However, it could be the case that inflation does
occur, but the Universe evolves towards homogeneity and isotropy starting from
a more complicated metric. There have been attempts to consider an entirely
arbitrary metric, meaning both inhomogeneous and anisotropic
\cite{Goldwirth:1989pr}; however, the calculations become cumbersome. In this
paper, we focus on homogeneous and anisotropic cosmology to extract analytical
information. This class of geometries \cite{Misner1973} exhibits exciting
cosmological features in the inflationary and post-inflationary epochs
\cite{peebles1993principles}. Nine spatially homogeneous, but generally anisotropic, Bianchi models exist based on
the real three-dimensional Lie algebra classification. In these spacetimes,
three-dimensional hypersurfaces are defined by the orbits of three isometries.
An essential characteristic of the Bianchi models is that the physical
variables depend only on time, which means that the field equations are a
system of ordinary differential equations \cite{mx2,mx3}. In recent years, the
class of anisotropic geometries has gained much interest because of anisotropic
anomalies in the CMB and Large-Scale Structure
(LSS) data. The origin of asymmetry and other measures of statistical
anisotropy on the large scales of the Universe is a long-standing open
question in cosmology. ``Planck Legacy'' temperature anisotropy data
\cite{Planck:2018nkj} show strong evidence of violating the Cosmological
Principle in its isotropic aspect \cite{Fosalba:2020gls, LeDelliou:2020kbm}.

The family of spatially homogeneous Bianchi cosmologies includes as subclasses
many important gravitational models, such as the Mixmaster Universe or the
isotropic FLRW spacetimes \cite{mr,Mis69,mis1,mx1}. As expected, the latter
comes as a limit of Bianchi models where the anisotropy vanishes. Indeed, the
flat, the open, and the closed FLRW geometries are related to the Bianchi I,
V and IX spacetimes respectively \cite{WE}. In general, the Bianchi
spacetimes are defined by three scale factors \cite{mr}; however, the locally
rotational spacetimes (LRS) admit an extra fourth isometry, and the LRS
Bianchi line elements admit two independent scale factors. It is interesting
to mention that the LRS Bianchi IX spacetime is related to the Kantowski-Sachs
geometry \cite{ks1}.

Symmetric Teleparallel General Relativity (STGR) \cite{ngr2,lav1} is a
geometric theory of gravity, equivalent to General Relativity. The theory is
described by a metric tensor $g_{\mu\nu}\left(  x^{\lambda}\right)  $ and a
symmetric flat connection $\Gamma_{\;\lambda\nu}^{\kappa}\left(  x^{\lambda
}\right)  $, different from that of the Levi-Civita connection. It is assumed
that the connection $\Gamma_{\;\lambda\nu}^{\kappa}\left(  x^{\lambda}\right)
$ admits the same symmetries with the metric tensor $g_{\mu\nu}\left(
x^{\lambda}\right)  $ and the corresponding Riemann tensor is assumed to have
zero components (flat connection). Since the connection is considered to be
symmetric, it means that its torsion also vanishes. The nonmetricity tensor
plays an important role in symmetric teleparallel geometries, since it is the
fundamental geometric object used to define the gravitational Action Integral
\cite{lav1} and thus describe gravitational interactions.

To explain the observed cosmic acceleration \cite{rr1,Teg,Komatsu,suzuki11} in
the context of symmetric teleparallel theory in a natural way, it has been
proposed the employ of nonlinear terms of the nonmetricity scalar, $Q$, in
the gravitational action, leading to the symmetric teleparallel $f\left(
Q\right) $-theory \cite{lav2,lav3}. The same approach has been considered
before in the case of $f\left(  R\right)  ~$\cite{fr} and $f\left(  T\right)
$ theories of gravity \cite{ft}, where the Ricci scalar $R$ of the Levi-Civita
connection and the torsion scalar $T$ of the teleparallel connection
\cite{Weitzenb23} are considered to define the gravitational theory. For a
comprehensive review on teleparallel theories of gravity see
\cite{Bahamonde:2021gfp}. The novelty of the $f\left(  X\right)  $-theories
\cite{rev1,rev2}, where $X$ is a geometric scalar, is that new terms which are
introduced in the modified field equations drive the dynamics to describe the
expansion of the universe in a geometric way, without the addition of any
exotic form of matter/energy \cite{cp1}.

Preliminary cosmological studies in $f\left(  Q\right)  $-theory have shown
that it is a potential geometric dark energy candidate which can challenge the
concordance model in cosmology \cite{ww0,ww8,ds00}. Numerous studies show that
$f\left(  Q\right)  $-theory can reproduce various cosmological scenarios
\cite{lym1,lym2,lym3,lym4}; in \cite{nd1}, the authors determined the criteria
for the existence of scaling cosmological solutions; the detailed phase-space
analysis was presented in \cite{andyn1} and it indicates that $f\left(
Q\right)  $-theory can be used to describe not only late-time but also
early-time acceleration phases of the Universe. Similar results are presented
in \cite{dyn2}. The Hamiltonian analysis of $f(Q)$-theory was studied in
\cite{hd1,hd2,hd3}, while quantum cosmology was investigated in
\cite{qdim1,qdim2}. For extensions of $f\left(  Q\right)  $-theory we refer
the reader to \cite{ex1,ex2,ex3,ex7} and other cosmological applications are
discussed in \cite{ex4,ex5,ex6} and references therein. A recent review on the
topic is \cite{Heisenberg:2023lru}.

Because the connection in $f\left(  Q\right)  $-theory is flat and symmetric,
it comes naturally that there exists a coordinate system where all the
components of the connection can vanish; as a result, the covariant
derivatives are reduced to partial derivatives. This characteristic coordinate
system is known as the coincident gauge \cite{lav3}. Different connections, in general, affect the dynamics in a given $f(Q)$-theory. However, in the case where $f(Q)$ is a linear function, the connection becomes dynamically irrelevant and consequently, there are no differences produced by the choice of the latter. The resulting differences in the case of a non-linear theory can be easily observed in studies of
isotropic and homogeneous cosmology \cite{Hohmann, Heis2, Zhao} and in static
spherical symmetric spacetimes \cite{seb1}. When the coincident gauge is
employed, the corresponding field equations of $f\left(  Q\right)  $-theory
are of second order involving only the metric degrees of freedom, while, when we assume an arbitrary connection, the
resulting field equations in the scalar-tensor representation are described by invoking 
two additional scalar fields \cite{mini1,lav3}. Although the coincident gauge can always
be recovered for a given geometry, it cannot always be used blindly, when a specific assumption for the spacetime metric has been made. For instance, for a homogeneous and isotropic FLRW
metric with non-zero spatial curvature, written in a diagonal form, and in the usual spherical coordinate system, the unique connection which can
produce the field equations for a nonlinear function $f\left(  Q\right)  $ is
defined in a purely non-coincident gauge \cite{Zhao}. The same property is true
for the case of anisotropic spacetimes with curvature \cite{dimks} and
non-flat inhomogeneous geometries \cite{seb1}. This happens because assuming a specific form for the spacetime metric already fixes partially the gauge, and this choice may be incompatible with the coordinate system where the connection components vanish.

The structure of the paper is
as follows. In Section \ref{sec2} we briefly discuss the basic properties and definitions
of symmetric teleparallel general relativity and of the symmetric teleparallel
$f\left(  Q\right)  $-theory of gravity. We discuss previous results for the
spatially flat and isotropic universe in Section \ref{sec3}. Homogeneous and
isotropic locally rotational spacetimes with nonzero spatial curvature in
symmetric teleparallel $f\left(  Q\right)  $-theory are introduced in Section
\ref{sec4}. We give emphasis in the Kantowski-Sachs and Bianchi III geometries
and we present the gravitational field equations in the case of vacuum for a
nonlinear function $f\left(  Q\right)  $. Section \ref{sec5} includes the main
results of this analysis where we present a detailed analysis of the
phase-space for the anisotropic cosmological model. From our analysis it
follows that for specific values of the free parameters which define the
connection the theory can provide the limit of General Relativity (GR) and
there exist a plethora of asymptotic solutions which can describe anisotropic
inflationary solutions. However, in the generic case of the connection these
solutions are lost. Thus the cosmological viability of the theory constraints
the free parameters of the connection as it follows from the analysis of the
asymptotics. Finally, in Section \ref{sec6} we discuss our results.

\section{Symmetric Teleparallel Geometry and gravity}

\label{sec2}

In teleparallel theories, parallelism at a distance is achieved by the
vanishing of the curvature of the connection, i.e. $R^{\alpha}{}_{\mu\beta\nu}
= 0$, which makes the connection become integrable and thus it can be
expressed as
\begin{equation}
\Gamma^{\alpha}{}_{\mu\nu} = (\Lambda^{-1}) ^{\alpha}{}_{\lambda}\partial
_{\mu}\Lambda^{\lambda}{}_{\nu}\,,
\end{equation}
where $\Lambda\in GL(4,\mathbf{R})$. In addition, symmetric means that the
torsion of the connection vanishes, i.e. $T^{\alpha}{}_{\mu\nu} =
2\Gamma^{\alpha}{}_{[\mu\nu]} = 0$, in which case $\Lambda$ can be written as
$\Lambda^{\alpha}{}_{\beta}= \partial_{\beta}\xi^{\alpha}$, with $\xi^{\alpha
}$ being an arbitrary coordinate system. This leads to the symmetric
teleparallel connection that can be expressed as
\begin{equation}
\label{sym-tel-con}\Gamma^{\alpha}{}_{\mu\nu} = \frac{\partial x^{\alpha}%
}{\partial\xi^{\lambda}} \partial_{\mu}\partial_{\nu}\xi^{\lambda}\,.
\end{equation}
Since $\xi^{\alpha}$ is arbitrary, we can always find a coordinate system in
which the connection vanishes by performing a diffeomorphism; this is called
the coincident gauge.

Notice that, the theory of gravity could be perfectly formulated just by the
metric tensor, $g_{\mu\nu}$, with the kinetic term in the action being
$\partial_{\alpha}g_{\mu\nu}$. However, this would not have the same
symmetries as General Relativity (GR); it would violate diffeomorphism (Diff)
invariance. In order to resolve that, we can employ the above $\xi$'s
as St\"uckelberg fields and the Diff symmetry will be restored.

According to the above, the only non-trivial geometric object in a symmetric
teleparallel geometry is the nonmetricity tensor, expressed as
\begin{equation}
Q_{\alpha\mu\nu}=\partial_{\alpha}g_{\mu\nu}-\Gamma^{\lambda}{}_{\alpha\mu
}g_{\lambda\nu}-\Gamma^{\lambda}{}_{\alpha\nu}g_{\lambda\mu}=\partial_{\alpha
}g_{\mu\nu}-2\frac{\partial x^{\sigma}}{\partial\xi^{\lambda}}\partial
_{\alpha}\partial_{(\mu}\xi^{\lambda}g_{\nu)\sigma}\,.
\end{equation}
This object transforms clearly covariantly and thus any theory formulated with
it will be automatically Diff invariant.

\subsection{Symmetric Teleparallel Equivalent of General Relativity}

By defining the two independent traces $Q_{\mu}= Q_{\mu\nu}{}^{\nu}$ and
$\tilde{Q}^{\nu}{}_{\mu\nu}$, the nonmetricity scalar is defined as
\begin{equation}
\label{non-metricity-scalar}Q = Q_{\alpha\mu\nu}P^{\alpha\mu\nu}\,,
\end{equation}
where
\begin{equation}
P^{\alpha}{}_{\mu\nu} = -\frac{1}{4}Q^{\alpha}{}_{\mu\nu} + \frac{1}{2}
Q_{(\mu} {}^{\alpha}{}_{\nu)} + \frac{1}{4}g_{\mu\nu}(Q^{\alpha}- \tilde
{Q}^{\alpha}) - \frac{1}{4}\delta^{\alpha}{}_{(\mu} Q_{\nu)}\,,
\end{equation}
with $\delta{\alpha}_{\mu}$ being the 4-dimensional Kronecker delta and the
brackets denote symmetrization $2A_{(\mu\nu)} = A_{\mu\nu} + A_{\nu\mu}$.

As mentioned above, the curvature of the symmetric teleparallel connection
\eqref{sym-tel-con} is zero. However, the curvature calculated from the
Levi-Civita connection, $\accentset{\circ}{R}^{\alpha}{}_{\mu\beta\nu}$, is
not. The relation between the nonmetricity scalar $Q$ and the Ricci scalar
$\accentset{\circ}{R}$ is given by
\begin{equation}
\accentset{\circ}{R} = Q + \accentset{\circ}{\nabla}_{\alpha}(Q^{\alpha}-
\tilde{Q}^{\alpha})\,.
\end{equation}
By taking the functional integral of the above, the last term will act as a
boundary term and thus contribute nothing at the dynamics of the theory. This
means that GR and STGR are two dynamically equivalent theories since
\begin{equation}
\mathcal{S}_{\mathrm{GR}} = \int\mathrm{d}^{4} x\sqrt{-g} \accentset{\circ}{R}
\sim\mathcal{S}_{\mathrm{STGR}} = \int\mathrm{d}^{4} x \sqrt{-g} Q \,.
\end{equation}
What is more, any prediction of GR should be predicted by STGR as well and any
solution in GR should have an analog solution in STGR. The only thing that
changes is the geometric interpretation of gravitational interactions.

\subsection{$f(Q)$-Theory}

In the same spirit as with $f(\accentset{\circ}{R})$-theories, we can
generalize the nonmetricity scalar \eqref{non-metricity-scalar} with a
general function of it, so that the new action will read
\begin{equation}
\label{f(Q)-action}\mathcal{S}_{f(Q)} = \int\mathrm{d} ^{4}x \sqrt{-g} f(Q)\,.
\end{equation}
Since the action contains non-linear terms in $Q$, not only the two theories
are no longer equivalent, but also solutions, like the non-flat FLRW one,
which worked fine in the coincident gauge STGR, are no longer solutions in
$f(Q)$.

So in this case, we have both the metric and the connection
\eqref{sym-tel-con} as fundamental variables. Varying the action
\eqref{f(Q)-action} with respect to the metric, we get
\begin{equation}
\frac{2}{\sqrt{-g}} \nabla_{\lambda}(\sqrt{-g}f^{\prime}(Q)P^{\lambda}_{\;\;\mu\nu}) -
\frac{1}{2}f(Q)g_{\mu\nu} + f^{\prime}(Q)(P_{\mu\rho\sigma} Q_{\nu}{}%
^{\rho\sigma} - 2 Q_{\rho\sigma\mu}P^{\rho\sigma}{}_{\nu}) = 0\,,
\end{equation}
where the prime denotes differentiation with respect to the argument, i.e.
$f^{\prime}(Q) = f_{,Q}$. Respectively, varying the action with respect to the
connection\footnote{Obviously, instead of the connection $\Gamma^{\alpha}%
{}_{\mu\nu}$, one could vary the action with respect to the arbitrary $\xi$'s;
the equations of motion would be the same.} we get
\begin{equation}
\nabla_{\mu}\nabla_{\nu}(\sqrt{-g}f^{\prime}(Q)P^{\mu\nu}_{\quad\sigma}) = 0\,.
\end{equation}
In the coincident gauge, the latter is satisfied identically.

\section{Flat FLRW Cosmology in $f(Q)$ gravity}

\label{sec3}

Consider a homogeneous and isotropic flat FLRW metric of the form
\begin{equation}
\mathrm{d}s^2=-N(t)^{2} \mathrm{d}t^2+a(t)^{2}\left(  \mathrm{d}r^{2}+r^{2}(\mathrm{d}\theta
^{2}+\sin^{2}\theta\mathrm{d}\phi^{2})\right)  \,,
\end{equation}
with $N$ being the lapse function and $a$ the scale factor. By forcing the
symmetries of this metric, i.e. three rotations and three translations, to an
arbitrary connection, we find that the connection depends on five unknown
functions (out of 64 independent components). Once we impose the symmetric
teleparallel constraints, i.e. vanishing curvature and torsion of the
connection, we end up with three unknown functions $\{C_{1},C_{2},C_{3}\}$ and
three constrain equations
\begin{gather}
C_{1}C_{3}-C_{3}^{2}-\dot{C}_{3}=0\,,\\
C_{1}C_{2}-C_{2}C_{3}+\dot{C}_{2}=0\,,\\
C_{2}C_{3}=0\,.
\end{gather}
Because of the last of the above, we have three cases

\begin{itemize}

\item {Case I: $C_{2} = 0 = C_{3}$}  where $\Gamma^{t} {}_{tt} = \gamma$.

\item {Case II: $C_{2} = 0$ and $C_{3} \neq0$}  where $\Gamma^{t} {}_{tt} =
\gamma+ \frac{\dot{\gamma} }{\gamma}$ and $\Gamma^{r}{}_{tr} =\Gamma^{r}%
{}_{rt} =\Gamma^{\theta}{}_{t\theta}=\Gamma^{\theta}{}_{\theta t}=\Gamma
^{\phi}{}_{t\phi}=\Gamma^{\phi}{}_{\phi t}=\gamma$.

\item {Case III: $C_{2} \neq0$ and $C_{3} = 0$}  where $\Gamma^{t} {}_{tt} = -
\frac{\dot{\gamma} }{\gamma}$ and $\Gamma^{t}{}_{rr} =\gamma, \Gamma^{t}
{}_{\theta\theta} = \gamma r^{2}, \Gamma^{t}{}_{\phi\phi} = \gamma r^{2}
\sin^{2} \theta$.
\end{itemize}

The rest of the components of the connection are the same as the Levi-Civita
one for the three-dimensional flat space, i.e.
\begin{gather}
\Gamma^{r}{}_{\theta\theta}=-r,\,\Gamma^{r}{}_{\phi\phi}=-r\sin^{2}%
\theta,\,\Gamma^{\theta}{}_{\phi\phi}=-\sin\theta\cos\theta,\,\Gamma^{\phi}%
{}_{\theta\phi}=\Gamma^{\phi}{}_{\phi\theta}=\cot\theta,\,\nonumber\\
\Gamma^{\theta}%
{}_{r\theta}=\Gamma^{\theta}{}_{\theta r}=\Gamma^{\phi}{}_{r\phi}=\Gamma
^{\phi}{}_{\phi r}=r^{-1}\,.
\end{gather}
Summarizing, a connection which is spatially flat, homogeneous, isotropic,
torsionless and with no curvature can be parametrized in the above three
distinct ways, which could lead to interesting phenomenology in cosmology.

\section{Kantowski-Sachs and Bianchi III geometry}

\label{sec4}

We proceed our study by considering the anisotropic cosmological models with
line element
\begin{equation}
ds^{2}=-N(t)^{2}dt^{2}+a^{2}\left(  t\right)  \left(  e^{2b\left(  t\right)
}dx^{2}+e^{-b\left(  t\right)  }\left(  dy^{2}+S^{2}\left(  y\right)
dz^{2}\right)  \right)  , \label{ks.01}%
\end{equation}
with $S\left(  y\right)  =\sin y$ or $S\left(  y\right)  =\sinh y$.

For $S(y)=\sin y$ the line element\ (\ref{ks.01}) describes the
Kantowski-Sachs space, while for $S(y)=\sinh y$ the line element\ (\ref{ks.01}%
) corresponds to the locally rotational (LRS) Bianchi III geometry. The scale
factor $a\left(  t\right)  $ describes the size of the universe; that is, the
volume of the three-dimensional space is defined as $V=a^{3}$. Moreover,
$b\left(  t\right)  $ is the anisotropic parameter.

Indeed, spacetime (\ref{ks.01}) admits a four-dimensional Lie algebra
consisted by the vector fields%

\[
\xi_{1}=\partial_{z}~,~\xi_{2}=\cos z\, \partial_{y}-\frac{S^{\prime}%
(y)}{S(y)}\sin z \, \partial_{z}%
\]

\[
\xi_{3}=\sin z\,\partial_{y}+\frac{S^{\prime}(y)}{S(y)}\cos z\,\partial
_{z}\text{ and }\xi_{4}=\partial_{x}.
\]

In this study we consider the symmetric and flat connection $\Gamma_{\kappa
\nu}^{\mu}$ with nonzero components \cite{dimks}%

\begin{equation}%
\begin{split}
&  \Gamma_{\;tt}^{t}=-\frac{1}{\gamma_{2}}\left[  \dot{\gamma}_{2}+c_{1}%
\gamma_{1}\left(  2-c_{2}\gamma_{1}\right)  +k\right]  ,\quad\Gamma_{\;tx}%
^{t}=c_{1}\left(  1-c_{2}\gamma_{1}\right)  ,\quad\Gamma_{\;xx}^{t}=c_{1}%
c_{2}\gamma_{2},\\
&  \Gamma_{\;yy}^{t}=\gamma_{2},\quad\Gamma_{\;zz}^{t}=\gamma_{2}%
S(y)^{2},\quad\Gamma_{\;tt}^{x}=\frac{1}{\gamma_{2}^{2}}\left[  \gamma
_{1}\left(  k+c_{1}\gamma_{1}\right)  \left(  c_{2}\gamma_{1}-1\right)
-\gamma_{2}\dot{\gamma}_{1}\right] \\
&  \Gamma_{\;tx}^{x}=-\frac{c_{2}\gamma_{1}}{\gamma_{2}}\left(  k+c_{1}%
\gamma_{1}\right)  ,\quad\Gamma_{\;xx}^{x}=c_{1}+c_{2}k+c_{1}c_{2}\gamma
_{1},\quad\Gamma_{\;yy}^{x}=\gamma_{1},\quad\Gamma_{\;zz}^{x}=\gamma
_{1}S(y)^{2},\\
&  \Gamma_{\;ty}^{y}=\Gamma_{\;tz}^{z}=-\frac{k+c_{1}\gamma_{1}}{\gamma_{2}%
},\quad\Gamma_{\;xy}^{y}=\Gamma_{\;xz}^{z}=c_{1},\quad\Gamma_{\;zz}%
^{y}=-S(y)S^{\prime}(y),\quad\Gamma_{\;yz}^{z}=\frac{S^{\prime}(y)}{S(y)},
\end{split}
\end{equation}
with $\gamma_{1}=-\frac{1}{c_{2}}-\frac{k}{c_{1}}$. This choice is needed in order to eliminate the non-diagonal term produced in the field equations. The $k$ in the previous relations is a constant given by $k=-S''(y)/S(y)$ and it can be equal to $+1$ or $-1$.

For the latter connection and the line element (\ref{ks.01}) the nonmetricity
scalar is derived to be
\begin{align}
Q  &  =-6H^{2}+2\frac{k}{a^{2}}e^{b}+\frac{3}{2}\frac{\dot{b}^{2}}{N^{2}%
}-\frac{6kH}{\gamma_{2}N}+\frac{e^{b}}{a^{2}}\left(  2+c_{1}c_{2}%
e^{-3b}\right)  \left(  HN\gamma_{2}+\dot{\gamma}_{2}+\frac{\gamma_{2}\dot{N}%
}{N}\right) \nonumber\\
&  +2\frac{e^{b}}{a^{2}}\left(  1-c_{1}c_{2}e^{-3b}\right)  \gamma_{2}\dot
{b}+\frac{3(2c_{1}+c_{2}k)^{2}}{4c_{1}c_{2}}\frac{H}{N\gamma_{2}}-\frac
{(c_{2}k-2c_{1})^{2}}{4c_{1}c_{2}N^{2}\gamma_{2}}\left(  \frac{\dot{N}}%
{N}+\frac{\dot{\gamma_{2}}}{\gamma_{2}}\right) , \label{qq.01}%
\end{align}
in which $H$ stands for $H=\frac{\dot{a}}{N a}$.

Hence, the gravitational field equations in the vacuum are \cite{dimks}

$tt:$%
\begin{equation}%
\begin{split}
&  f^{\prime}(Q)\left(  3H^{2}+\frac{k}{a^{2}}e^{b}-\frac{3}{4}\frac{\dot
{b}^{2}}{N^{2}}\right)  +\frac{1}{2}\left(  f(Q)-Qf^{\prime}(Q)\right) \\
&  +\frac{\dot{Q}}{N}f^{\prime\prime}(Q) \left[  \frac{(c_{2} k-2 c_{1})^{2}%
}{8 c_{1} c_{2} N \gamma_{2}} - \frac{ e^{b} N \gamma_{2}}{a^{2}}\left(
1+\frac{c_{1} c_{2}}{2} e^{-3 b}\right)  \right]  =0, \label{qq.02}%
\end{split}
\end{equation}

$xx:$%
\begin{equation}%
\begin{split}
&  f^{\prime}(Q) \left(  \frac{\ddot{b}}{N^{2}} +3H\frac{\dot{b}}{N} -
\frac{\dot{b}\dot{N}}{N^{3}} -2\frac{\dot{H}}{N}-3H^{2}-\frac{k e^{b}}{a^{2}%
}-\frac{3}{4}\frac{\dot{b}^{2}}{N^{2}} \right)  -\frac{1}{2}\left(
f(Q)-Qf^{\prime}(Q)\right) \\
&  + \frac{\dot{Q}}{N}f^{\prime\prime}(Q) \left[  \frac{(c_{2} k-2 c_{1})^{2}%
}{8 c_{1} c_{2} N \gamma_{2}} + \frac{ e^{b} N \gamma_{2}}{a^{2}}\left(  1 -
\frac{c_{1} c_{2}}{2} e^{-3 b}\right)  + \frac{\dot{b}}{N} -2 H \right]
=0, \label{qq.03}%
\end{split}
\end{equation}

$yy,zz:$%
\begin{equation}%
\begin{split}
&  f^{\prime}(Q)\left(  \frac{\ddot{b}}{N^{2}}+3H\frac{\dot{b}}{N}-\frac
{\dot{b}\dot{N}}{N^{3}}+4\frac{\dot{H}}{N}+6H^{2}+\frac{3}{2}\frac{\dot{b}%
^{2}}{N^{2}}\right)  +\left(  f(Q)-Qf^{\prime}(Q)\right) \\
&  +\frac{\dot{Q}}{N}f^{\prime\prime}(Q)\left[  4H+\frac{\dot{b}}{N}-
\frac{(c_{2}k-2c_{1})^{2}}{4c_{1}c_{2}N\gamma_{2}}-\frac{c_{1}c_{2}e^{-2b}%
}{a^{2}}N\gamma_{2}\right]  =0 .
\end{split}
\end{equation}

In order to write the field equations in a simpler form we introduce the
scalar field $\phi=f^{\prime}\left(  Q\right)  $, the potential function
$V\left(  \phi\right)  =\left(  f(Q)-Qf^{\prime}(Q)\right)  $, from where we
can write the point-like Lagrangian \cite{mini1}
\begin{equation}
\label{Lplike}
\begin{split}
L(a,\dot{a},b,\dot{b},\phi,\dot{\phi},\Psi,\dot{\Psi})  &
=\frac{1}{N}\left(\frac{3}{2}a^{3}\phi\dot{b}^{2}  -6a\phi\dot{a}^{2}
-\frac{a^{3}(k-2\alpha)^{2}\dot{\Psi}\dot{\phi}}{4\alpha}\right)\\
&  -N\left( \frac{ae^{-2b}\left(  2e^{3b}+\beta\right)
\dot{\phi}}{\dot{\Psi}} -2kae^{b}\phi-a^{3}V(\phi)\right)  ,
\end{split}
\end{equation}
with $\gamma_{2}=\frac{1}{\dot{\Psi}}$,~$\alpha=\frac{c_{1}}{c_{2}}$ and
$\beta=c_{1}c_{2}$.

The field equations follow from the variation of the latter Lagrangian with
respect to the dynamical variables $\left\{  N,a,b,\phi,\Psi\right\}  $.
Specifically, the gravitational field equations are (here and henceforth we impose $N=1$ for the lapse function)

\begin{align}
\dot{H}= & \frac{12 a \dot{a} \dot{\phi}+6 \phi  \dot{a}^2-2 k e^{b} \phi }{12 a \phi }+\frac{a \left(6 \alpha  \phi  \dot{b}^2+(k-2 \alpha )^2 \dot{\Psi} \left(-\dot{\phi}\right)+4 \alpha  V(\phi )\right)}{16 \alpha  \phi }\nonumber\\ &-\frac{\left(\beta  e^{-2 b}+2
   e^{b}\right) \dot{\phi}}{12 a \phi  \dot{\Psi}}+\frac{a \left(6 \alpha  \phi  \dot{b}^2+(k-2 \alpha )^2 \dot{\Psi} \left(-\dot{\phi}\right)+4 \alpha  V(\phi )\right)}{16 \alpha  \phi },\label{f-eq1}\\
\ddot{b}= &  -\frac{3 \dot{a} \dot{b}}{a}+\frac{2 e^{-2 b} \left(k e^{3 b}+\frac{\left(\beta
   -e^{3 b}\right) \dot{\phi}}{\phi  \dot{\Psi}}\right)}{3 a^2}-\frac{\dot{b}
   \dot{\phi}}{\phi },\label{f-eq2}\\
   \ddot{\phi}= & \frac{\dot{\phi} \left(\dot{a} \left(40 \alpha  a^2 e^{2 b} (k-2 \alpha )^2
   \left(2 e^{3 b}+\beta \right) \dot{\Psi}^2-3 a^4 e^{4 b} (k-2 \alpha )^4
   \dot{\Psi}^4+16 \alpha ^2 \left(2 e^{3 b}+\beta \right)^2\right)\right)}{a \left(a^2 e^{2 b} (k-2 \alpha )^2 \dot{\Psi}^2-4 \alpha  \left(2 e^{3 b}+\beta \right)\right)^2}\nonumber\\ & +\frac{\dot{\phi} \left(8 \alpha  a
   \left(6 \alpha  a^2
   e^{2 b} \left(2 e^{3 b}+\beta \right) \dot{b}^2 \dot{\Psi}+4 \alpha  e^{2
   b} \left(2 e^{3 b}+\beta \right) \dot{\Psi} \left(a^2 V'(\phi )+2 k
   e^{b}\right)\right)\right)}{a \left(a^2 e^{2 b} (k-2 \alpha )^2 \dot{\Psi}^2-4 \alpha  \left(2 e^{3 b}+\beta \right)\right)^2}\nonumber\\ &+\frac{\dot{\phi} \left(8 \alpha  a
   \left(\left(e^{3 b}-\beta \right) \dot{b} \left(a^2 e^{2 b} (k-2 \alpha
   )^2 \dot{\Psi}^2+4 \alpha  \beta +8 \alpha  e^{3 b}\right)\right)\right)}{a \left(a^2 e^{2 b} (k-2 \alpha )^2 \dot{\Psi}^2-4 \alpha  \left(2 e^{3 b}+\beta \right)\right)^2}\nonumber\\&+\frac{\dot{\phi} \left( -192 \alpha ^2
   a e^{2 b} \dot{a}^2 \left(2 e^{3 b}+\beta \right) \dot{\Psi}\right)}{a \left(a^2 e^{2 b} (k-2 \alpha )^2 \dot{\Psi}^2-4 \alpha  \left(2 e^{3 b}+\beta \right)\right)^2},\label{f-eq3} \\
\ddot{\Psi}= & \frac{\dot{\Psi} \left(8 \alpha  e^{3 b} \left(\dot{a}+a \left(\dot{b}+k \Psi
   '\right)\right)+a e^{2 b} \dot{\Psi} \left(3 a (k-2 \alpha )^2 \dot{a}
   \dot{\Psi}-24 \alpha  \dot{a}^2+2 \alpha  a^2 \left(3 \dot{b}^2+2 V'(\phi
   )\right)\right)\right)}{4
   \alpha  a \left(2 e^{3 b}+\beta \right)-a^3 e^{2 b} (k-2 \alpha )^2
   \dot{\Psi}^2}\nonumber \\ &+\frac{\dot{\Psi} \left(4 \alpha  \beta  \left(\dot{a}-2 a \dot{b}\right)\right)}{4
   \alpha  a \left(2 e^{3 b}+\beta \right)-a^3 e^{2 b} (k-2 \alpha )^2
   \dot{\Psi}^2}\label{f-eq4} .
\end{align}
Additionally, the Friedmann equation reads
\begin{equation}
\frac{2 k e^{b} \phi }{a^2}-\frac{\beta  e^{-2 b} \dot{\phi}}{a^2 \dot{\Psi}}-\frac{2 e^{b} \dot{\phi}}{a^2 \dot{\Psi}}-\frac{3}{2} \phi 
   \dot{b}^2+6 H^2 \phi +\frac{k^2 \dot{\Psi} \dot{\phi}}{4 \alpha }-k \dot{\Psi}
   \dot{\phi}+\alpha  \dot{\Psi} \dot{\phi}+V(\phi )=0. \label{Friedmann-eq}%
\end{equation}

For simplicity, we kept the scale factor $a$ in the field equations, but in
the computations, we write everything in terms of the Hubble function, that is
$a=e^{\int H\,dt}.$ In the following we consider the power-law potential $V\left(  \phi\right)
=V_{0}\phi^{\lambda}$, which correspond to a power-law $f\left(  Q\right)  $
function.

\section{Phase-space analysis\newline}

\label{sec5}

In this section, we set $k=2\alpha$ in the point-like Lagrangian %
\eqref{Lplike} and obtain a simplified version of system \eqref{f-eq1}-%
\eqref{f-eq4}. This choice corresponds to having $\gamma_1=0$ in the components of the connection. The ensuing equations are
\begin{align}
\dot{H}= &  -\frac{\alpha e^{b}}{3a^{2}}+\frac{\beta e^{-2b}\dot{\phi}%
}{12a^{2}\phi\dot{\Psi}}+\frac{e^{b}\dot{\phi}}{6a^{2}\phi\dot{\Psi}}-\frac
{3}{8}\dot{b}^{2}-\frac{H\dot{\phi}}{\phi}-\frac{3}{2}H^{2}-\frac{V(\phi
)}{4\phi}\label{f-eq1-subcase},\\
\ddot{b}= &  \frac{4\alpha e^{b}}{3a^{2}}+\frac{1}{3a^{2}\phi\dot{\Psi}%
}\left(  2\beta e^{-2b}\dot{\phi}-2e^{b}\dot{\phi}\right)  -3H\dot{b}%
-\frac{\dot{b}\dot{\phi}}{\phi}\label{f-eq2-subcase},\\
\ddot{\phi}= &  \frac{1}{2e^{3b}+\beta}\left(  3a^{2}e^{2b}\dot{b}^{2}%
\dot{\Psi}\dot{\phi}+2a^{2}e^{2b}\dot{\Psi}\dot{\phi}V^{\prime}(\phi)+2e^{3b}\dot{b}%
\dot{\phi}\right)  -\frac{1}{2e^{3b}+\beta}\left(  12a^{2}e^{2b}H^{2}\dot
{\Psi}\dot{\phi}+2\beta\dot{b}\dot{\phi}\right)  \nonumber\\
&  +\frac{1}{2e^{3b}+\beta}\left(  2e^{3b}H\dot{\phi}+\beta H\dot{\phi
}+8\alpha e^{3b}\dot{\Psi}\dot{\phi}\right) \label{f-eq3-subcase}, \\
\ddot{\Psi}= &  \frac{1}{2e^{3b}+\beta}\left(  \frac{3a^{2}e^{2b}\dot{b}%
^{2}\dot{\Psi}^{2}}{2}+a^{2}e^{2b}\dot{\Psi}^{2}V^{\prime}(\phi)+2e^{3b}\dot{b}\dot{\Psi
}\right)  -\frac{1}{2e^{3b}+\beta}\left(  6a^{2}e^{2b}H^{2}\dot{\Psi}%
^{2}+2\beta\dot{b}\dot{\Psi}\right)  \nonumber\\
&  +\frac{1}{2e^{3b}+\beta}\left(  2e^{3b}H\dot{\Psi}+\beta H\dot{\Psi
}+4\alpha e^{3b}\dot{\Psi}^{2}\right) \label{f-eq4-subcase} .
\end{align}
In this case, the Friedmann equation becomes 
\begin{equation}
\frac{4\alpha e^{b}\phi}{a^{2}}-\frac{\beta e^{-2b}\dot{\phi}}{a^{2}\dot{\Psi
}}-\frac{2e^{b}\dot{\phi}}{a^{2}\dot{\Psi}}-\frac{3}{2}\phi\dot{b}%
^{2}+6H^{2}\phi+V(\phi)=0. \label{Friedmann-eq-special-case}%
\end{equation}

At this point, we define the dimensionless variables
\begin{align}
\Omega _{R}& =\frac{2\alpha e^{b-2\int H\,dt}}{3H^{2}},\quad y^{2}=\frac{%
V(\phi )}{6H^{2}\phi },\quad \Sigma ^{2}=\frac{\dot{b}^{2}}{4H^{2}}, \\
x& =\frac{\dot{\phi}}{H\phi },\quad Z=\frac{H}{\dot{\Psi}},\quad w=-\frac{%
\left( 2e^{3b}+\beta \right) \dot{\phi}e^{-2(b+\int H\,dt)}}{6H^{2}\phi \dot{%
\Psi}},
\end{align}%
and the new independent variable $\tau =\ln a$, such that $x^{\prime }=\frac{%
dx}{d\tau }$.

Using these variables, we can write the modified Friedmann's equation as
\begin{equation}
\Omega _{R}-\Sigma ^{2}+w+y^{2}+1=0,  \label{Friedmann-special-case}
\end{equation}%
and solve it for $w$ to reduce the dimension of the dynamical system, that
is \begin{equation}
\label{def-of-w}
    w=\Sigma ^{2}-y^{2}-\Omega _{R}-1.
\end{equation}

The field equations are reduced to the following system of first order differential equations
\begin{align}
\label{ds-1-special}\Omega _{R}^{\prime }=& 2\Omega _{R}\left( 2\Sigma ^{2}+\Sigma
+x+y^{2}\right) ,   \\
\label{ds-2-special}y^{\prime }=& \frac{1}{2}y\left( 4\Sigma ^{2}+\lambda x+x+2y^{2}+2\right) ,
\nonumber \\
\Sigma ^{\prime }=& \Sigma ^{\prime 2}(\Sigma +1)+\frac{3}{2}\Omega
_{R}\left( 2-\frac{xZ}{\alpha }\right) +(\Sigma +2)y^{2}, \\
\label{ds-3-special}x^{\prime }=& x\left( \frac{x\left( 2\alpha \left( \Sigma ^{2}+\lambda
y^{2}-1\right) +\Omega _{R}(2\alpha +3\Sigma Z)\right) }{\alpha (-\Sigma
^{2}+y^{2}+\Omega _{R}+1)}+2(\Sigma -1)^{2}+y^{2}\right) , \\
\label{ds-4-special}Z^{\prime }=& \frac{Z\left( 2\alpha (\Sigma -1)^{3}(\Sigma +1)-\alpha
y^{2}((\Sigma -4)\Sigma +(\lambda +1)x+\Omega _{R}+3)\right) }\nonumber \\ & +\frac{Z\left(-\Omega _{R}\left(
2\alpha (\Sigma -1)^{2}+x(2\alpha +3\Sigma Z)\right) -\alpha y^{4}\right) }{%
\alpha \left( -\Sigma ^{2}+y^{2}+\Omega _{R}+1\right) },
\end{align}

Each stationary point of the latter system describes an asymptotic solution
with deceleration parameter
\begin{equation}
q=-1-\frac{\dot{H}}{H^{2}}=2\Sigma ^{2}+x+y^{2}.  \label{deceleration-q}
\end{equation}

The initial assumption that $k=2\alpha $ means that the system, the critical points, and their stability depends only on two parameters $\alpha $ and $\lambda $. Since $k=\pm 1,$ this means that $\alpha =\pm \frac{1}{2}.$ 
The dynamical system has the following equilibrium
points $(\Omega _{R},y,\Sigma ,x,Z)$:

\begin{enumerate}
\item The family $L_{1}=(\Omega _{Rc},0,-1,-1,-2\alpha ),$ where $\Omega
_{Rc}\in \mathbb{R}$ is a free parameter. Hence, the asymptotic solution is
that of anisotropic or Kantowski-Sachs or Bianchi III universe. On the
surface $\Omega _{Rc}=0$, the Bianchi type I dynamics are recovered. The value of
the deceleration parameter is $q(L_{1})=1,$ this means that $L_{1}$ defines
a decelerated solution. The eigenvalues are $\left\{ 0,-2,2,6,\frac{%
5-\lambda }{2}\right\} .$ The family is normally hyperbolic the stability is
given by the nonzero eigenvalues, $L_{1}$ is a saddle. From (\ref%
{deceleration-q}) we caclulate $H\left( t\right) =\frac{1}{2t}$, that is, $%
a\left( t\right) =a_{0}\sqrt{t}$ and $\dot{b}^{2}=\frac{2}{t^{2}}$.

\item The family $L_{2}=(\Omega _{Rc},0,1,-3,-\frac{2\alpha}{3}),$ where $%
\Omega _{Rc}\in \mathbb{R}$ is a free parameter.\ The value of the
deceleration parameter is $q(L_{2})=1$, which means it has the same physical properties with point $L_{1} $. The eigenvalues are $%
\left\{ 0,-6,-6,6,-\frac{3}{2}(\lambda -1)\right\} ,$ as $L_{1},$ $L_{2}$ is
a saddle.

\item $P_{1}=(-\frac{2}{3},0,0,0,0),$ with eigenvalues $\left\{ -2,2,-1+i%
\sqrt{3},-1-i\sqrt{3},1\right\} .$ $P_{1}$ is a saddle. The value of the
deceleration parameter is $q(P_{1})=0.$ The asymptotic solution is isotropic
and with nonzero spatially curvature. That is the limit of Milne universe,
since $\dot{b}=0$, $a\left( t\right) =a_{0}t$ and $\Omega _{R}\neq 0$. It is
a solution which provides the limit of GR in the theory.

\item $P_{2}=(2\left( \sqrt{13}-4\right) ,0,\frac{1}{2}\left( 3-\sqrt{13}%
\right) ,\frac{1}{2}\left( 7\sqrt{13}-25\right) ,0).$ The eigenvalues are

\begin{enumerate}
\item for $\alpha =\frac{1}{2}$, {\tiny $\left\{ 3\left(3- \sqrt{13}\right)
,3\left( 3-\sqrt{13}\right) ,\frac{1}{2}\left( 9-3\sqrt{13}\pm i\sqrt{%
14\left( 37\sqrt{13}-133\right) }\right) ,\frac{1}{4}\left( 7\sqrt{13}%
\lambda -25\lambda -5\sqrt{13}+23\right) \right\} ,$ }

\item for $\alpha =-\frac{1}{2}$, {\tiny $\left\{ 3\left( 3-\sqrt{13}%
\right) ,3\left( 3-\sqrt{13}\right) ,\frac{1}{2}\left( 9-3\sqrt{13}\mp i%
\sqrt{14\left( 37\sqrt{13}-133\right) }\right) ,\frac{1}{4}\left( 7\sqrt{13}%
\lambda -25\lambda -5\sqrt{13}+23\right) \right\} .$ }
\end{enumerate}

The point is an attractor for $\lambda <\frac{5\sqrt{13}-23}{7\sqrt{13}-25}$
and a saddle in any other case, the stability does not change if $\alpha =%
\frac{1}{2}$ or $-\frac{1}{2}.$ The value of the deceleration parameter is $%
q(P_{2})=\frac{1}{2}\left( \sqrt{13}-3\right) \approx 0.3>0,$ this means
that $P_{2}$ describes a decelerated solution. Those asymptotic solutions belong to an anisotropic Bianchi III universe because $\Omega _{R}\left(
P_{2}\right) <0$.

\item $P_{3}=(-2\left( 4+\sqrt{13}\right) ,0,\frac{1}{2}\left( 3+\sqrt{13}%
\right) ,\frac{1}{2}\left( -25-7\sqrt{13}\right) ,0),$ the eigenvalues are

\begin{enumerate}
\item for $\alpha =\frac{1}{2},$ {\tiny $\left\{ 3\left( 3+\sqrt{13}\right)
,3\left( 3+\sqrt{13}\right) ,\frac{1}{2}\left( 9+3\sqrt{13}\mp \sqrt{%
14\left( 133+37\sqrt{13}\right) }\right) ,\frac{1}{4}\left( -7\sqrt{13}%
\lambda -25\lambda +5\sqrt{13}+23\right) \right\} $, }

\item for $\alpha =-\frac{1}{2},$ {\tiny $\left\{ 3\left( 3+\sqrt{13}\right)
,3\left( 3+\sqrt{13}\right) ,\frac{1}{2}\left( 9+3\sqrt{13}\pm \sqrt{%
14\left( 133+37\sqrt{13}\right) }\right) ,\frac{1}{4}\left( -7\sqrt{13}%
\lambda -25\lambda +5\sqrt{13}+23\right) \right\} $. }
\end{enumerate}

The point is a saddle for $\lambda<\frac{23+5\sqrt{13}}{25+7\sqrt{13}}$ or
$\lambda>\frac{23+5\sqrt{13}}{25+7\sqrt{13}},$ the stability does not change
if $\alpha=\frac{1}{2}$ or $-\frac{1}{2}.$ The value of the deceleration
parameter is $q(P_{2})=\frac{1}{2}\left(  -3-\sqrt{13}\right)  \approx
-3.3<-1,$ this means that $P_{3}$ describes an accelerated anisotropic
universe with negative curvature, that is, a Bianchi III geometry.

\item $P_{4}=(\Omega_{R},y,\Sigma,x,z)=(\frac{6\lambda+6}{5-7\lambda}%
,\frac{\sqrt{6}\sqrt{\lambda(\lambda+20)-17}}{7\lambda-5},-\frac{2(\lambda
-2)}{7\lambda-5},\frac{18}{5-7\lambda},0).$ This point exist for $\lambda
\neq5/7$ and $\lambda\leq-10-3\sqrt{13}$ or $\lambda\geq3\sqrt{13}-10.$ The
eigenvalues are $l_{1}=-\frac{12(\lambda-2)}{7\lambda-5},$ $l_{2}=f_{1}%
(\alpha,\lambda),l_{3}=f_{2}(\alpha,\lambda),l_{4}=f_{3}(\alpha,\lambda
),l_{5}=f_{4}(\alpha,\lambda),$ where $f_{i}$ are complicated expressions
depending on the parameters. Setting $\alpha=\pm\frac{1}{2}$ slightly changes
the eigenvalues but not the stability, see FIG.\ref{Fig-1}. The point is a
saddle. The value of the deceleration parameter is $q(P_{4})=\frac
{2(\lambda-2)}{7\lambda-5},$ $P_{4}$ describes an accelerated solution for
$\frac{5}{7}<\lambda<2$ and a de Sitter solution for $\lambda=1.$
\begin{figure}[ptb]
\centering
\includegraphics[width=0.4\textwidth]{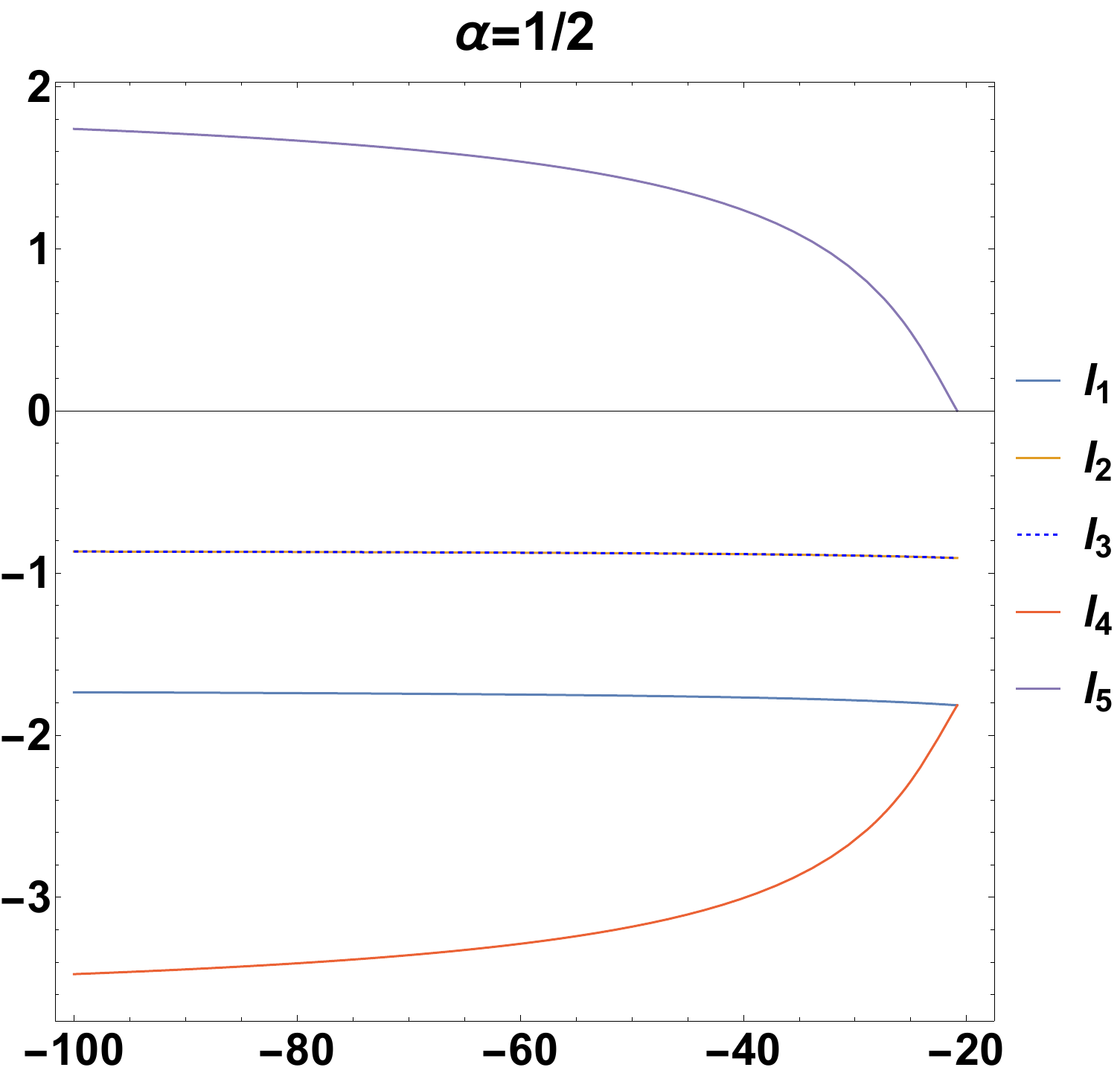}
\includegraphics[width=0.4\textwidth]{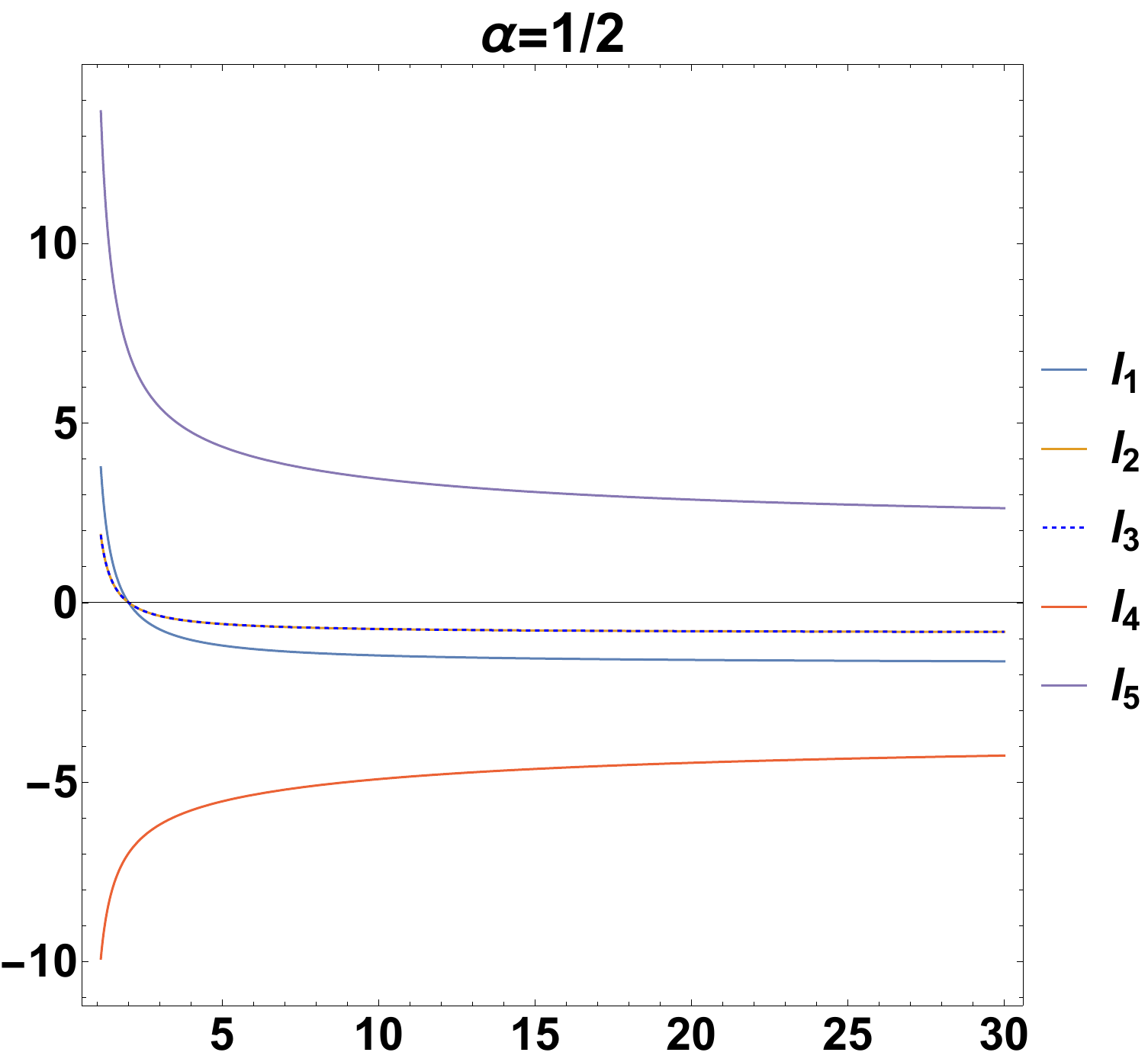}
\includegraphics[width=0.4\textwidth]{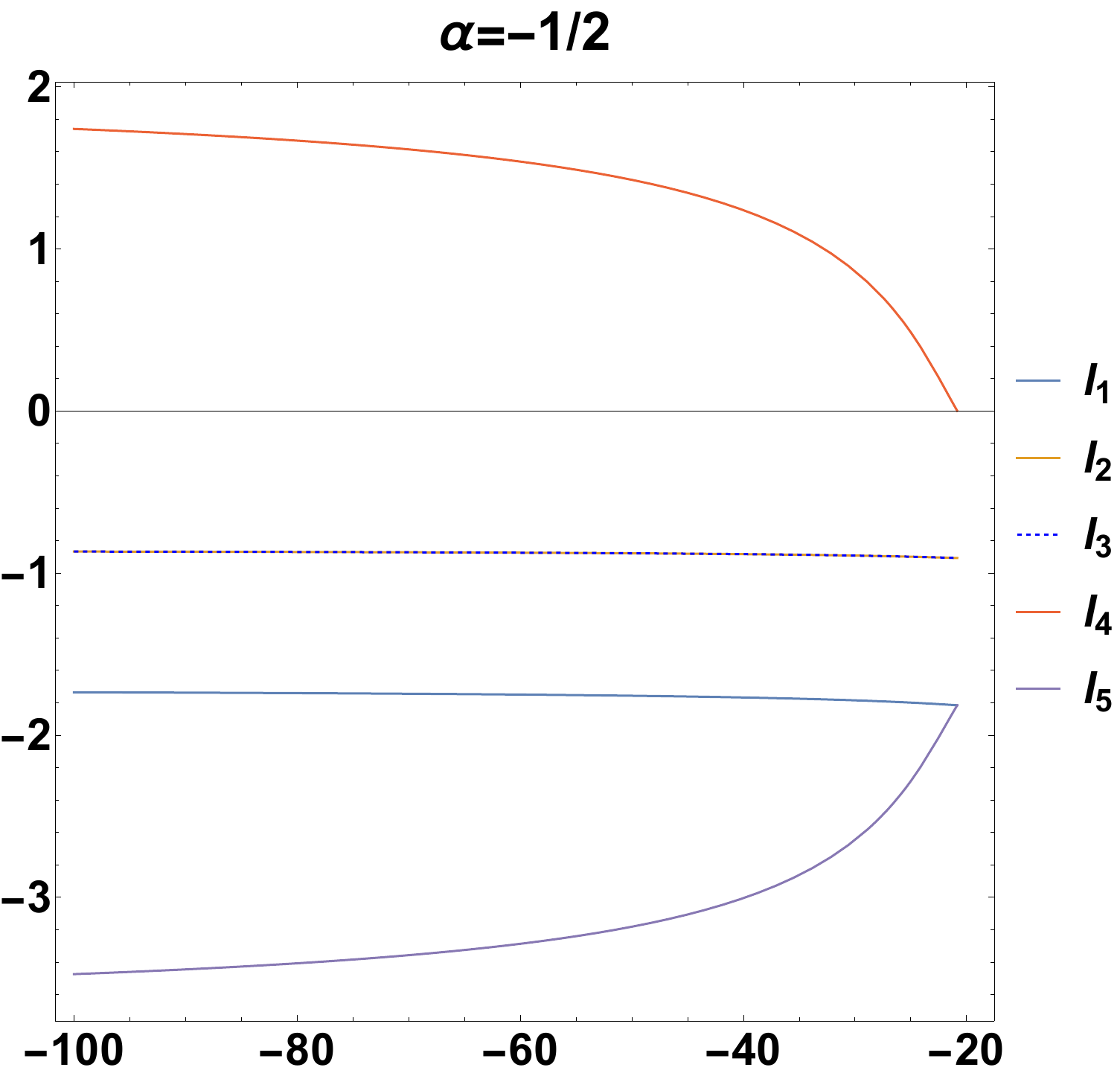}
\includegraphics[width=0.4\textwidth]{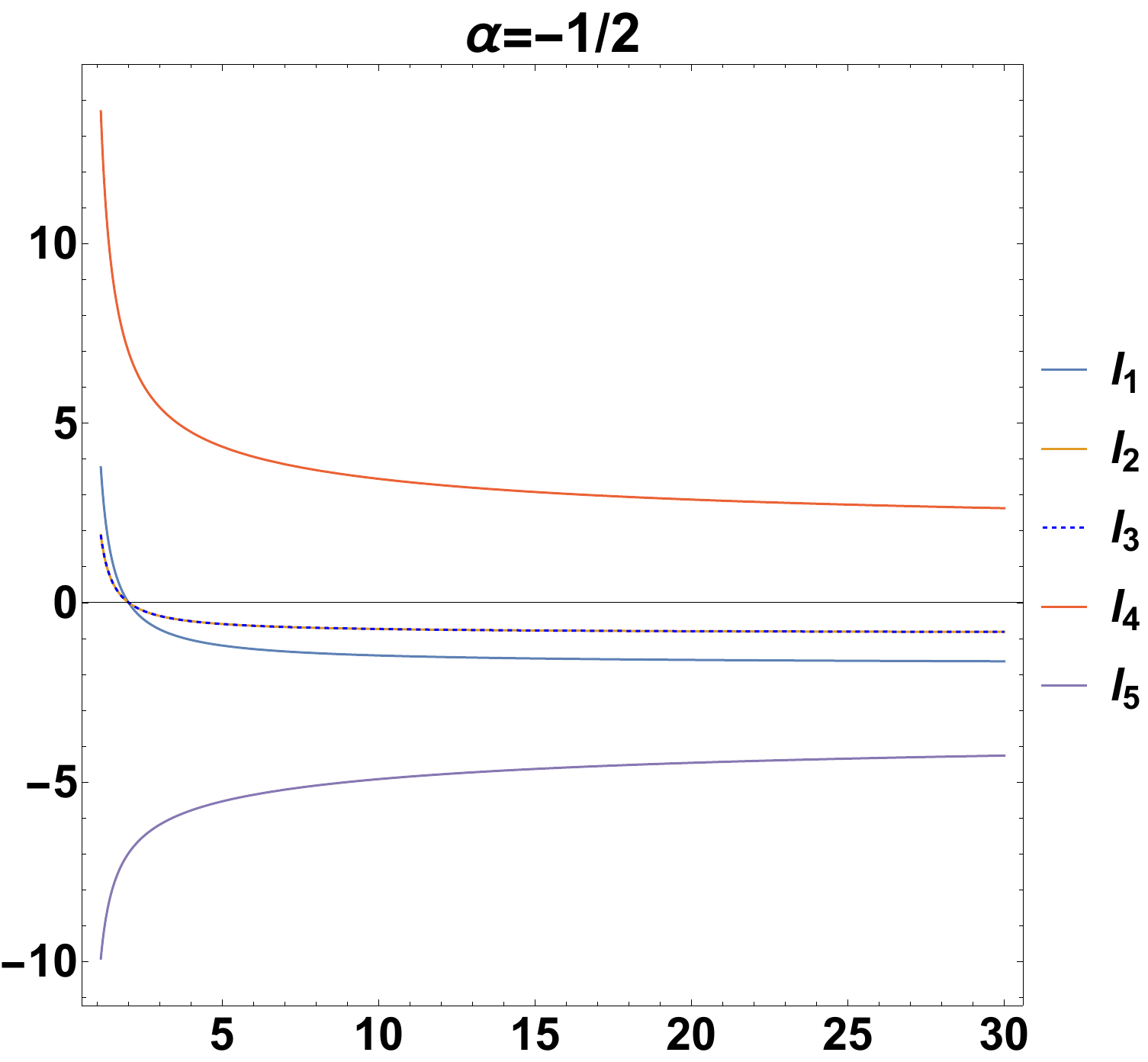}   \caption{Real part of the
eigenvalues for point $P_{4}$ for $\alpha=\pm\frac{1}{2}$ and different ranges
for $\lambda$ where the point exists.}%
\label{Fig-1}%
\end{figure}

\item $P_{5}=(\Omega_{R},y,\Sigma,x,z)=(\frac{6\lambda+6}{5-7\lambda}%
,\frac{\sqrt{6}\sqrt{\lambda(\lambda+20)-17}}{5-7\lambda},-\frac{2(\lambda
-2)}{7\lambda-5},\frac{18}{5-7\lambda},0).$ The difference with $P_{4}$ is a
minus sign in the $y$ coordinate, it exists for the same values of $P_{4}$ as
well. The stability is also the same (saddle) since they share eigenvalues see
FIG. \ref{Fig-1} for reference. The value of the deceleration parameter is
$q(P_{5})=\frac{2(\lambda-2)}{7\lambda-5},$ $P_{5}$ describes an accelerated
solution for $\frac{5}{7}<\lambda<2$ and a de Sitter solution for $\lambda=1.$
\end{enumerate}

\begin{figure}[ptb]
\centering
\includegraphics[width=0.4\textwidth]{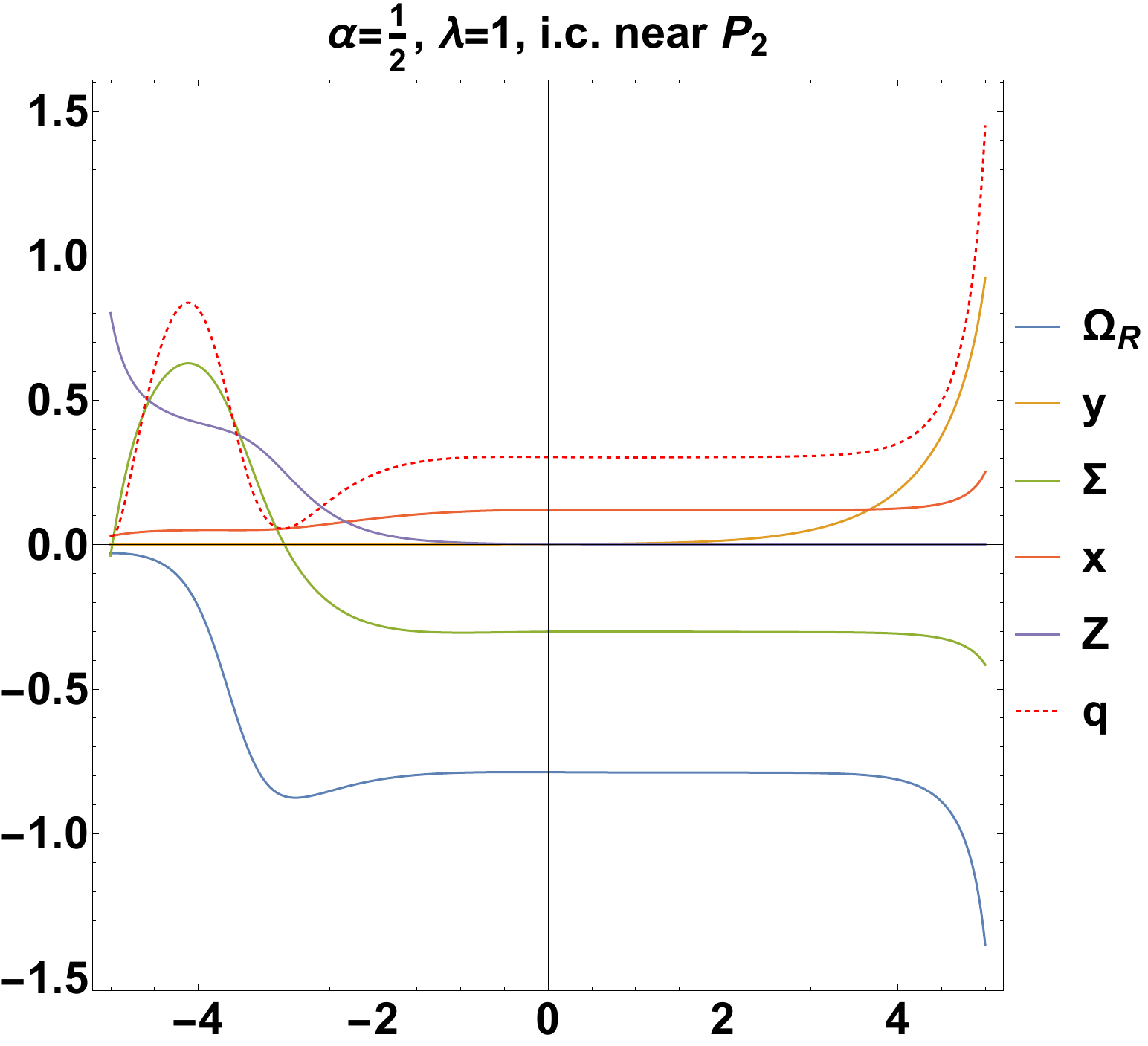}
\includegraphics[width=0.4\textwidth]{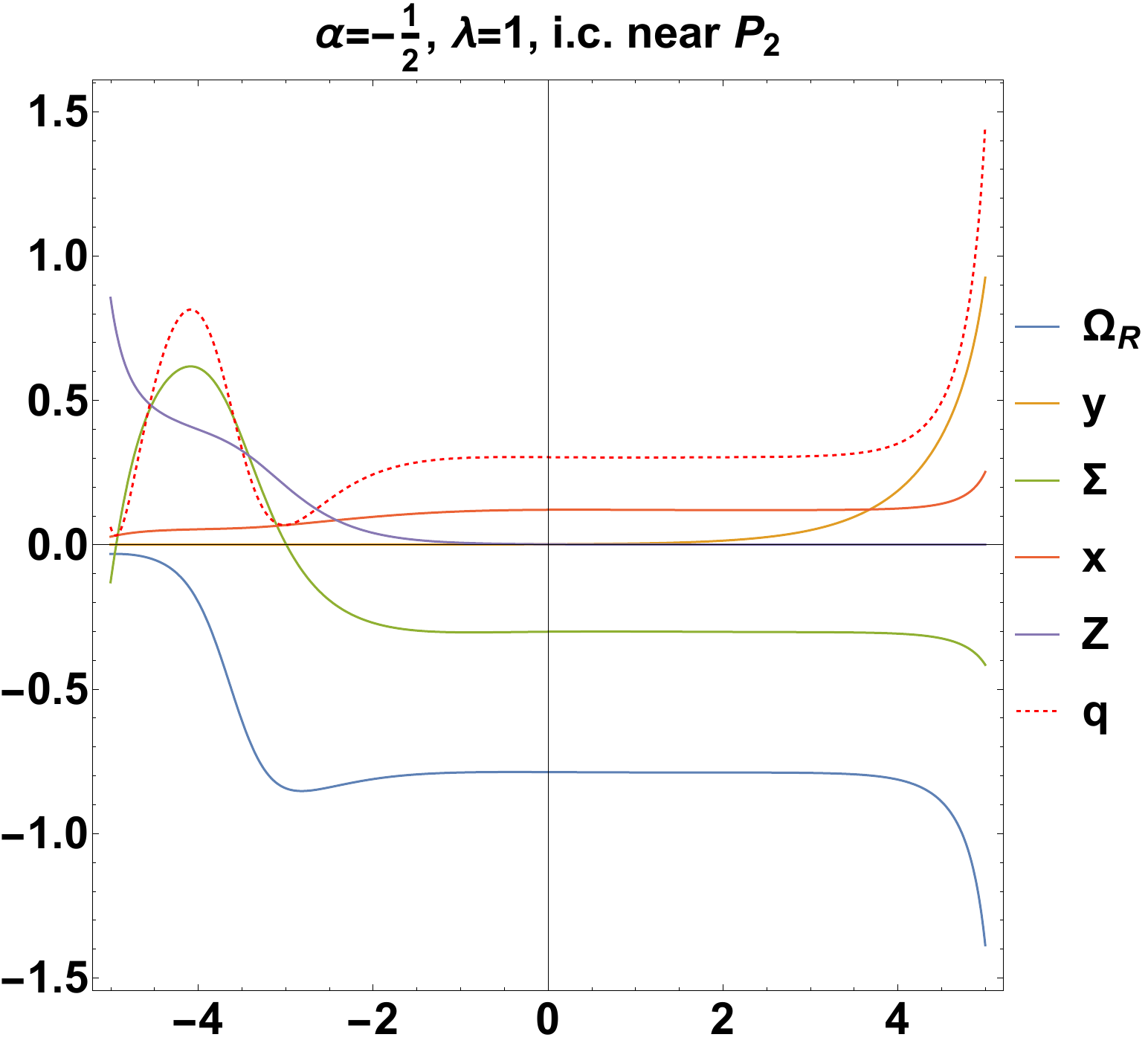}
\includegraphics[width=0.4\textwidth]{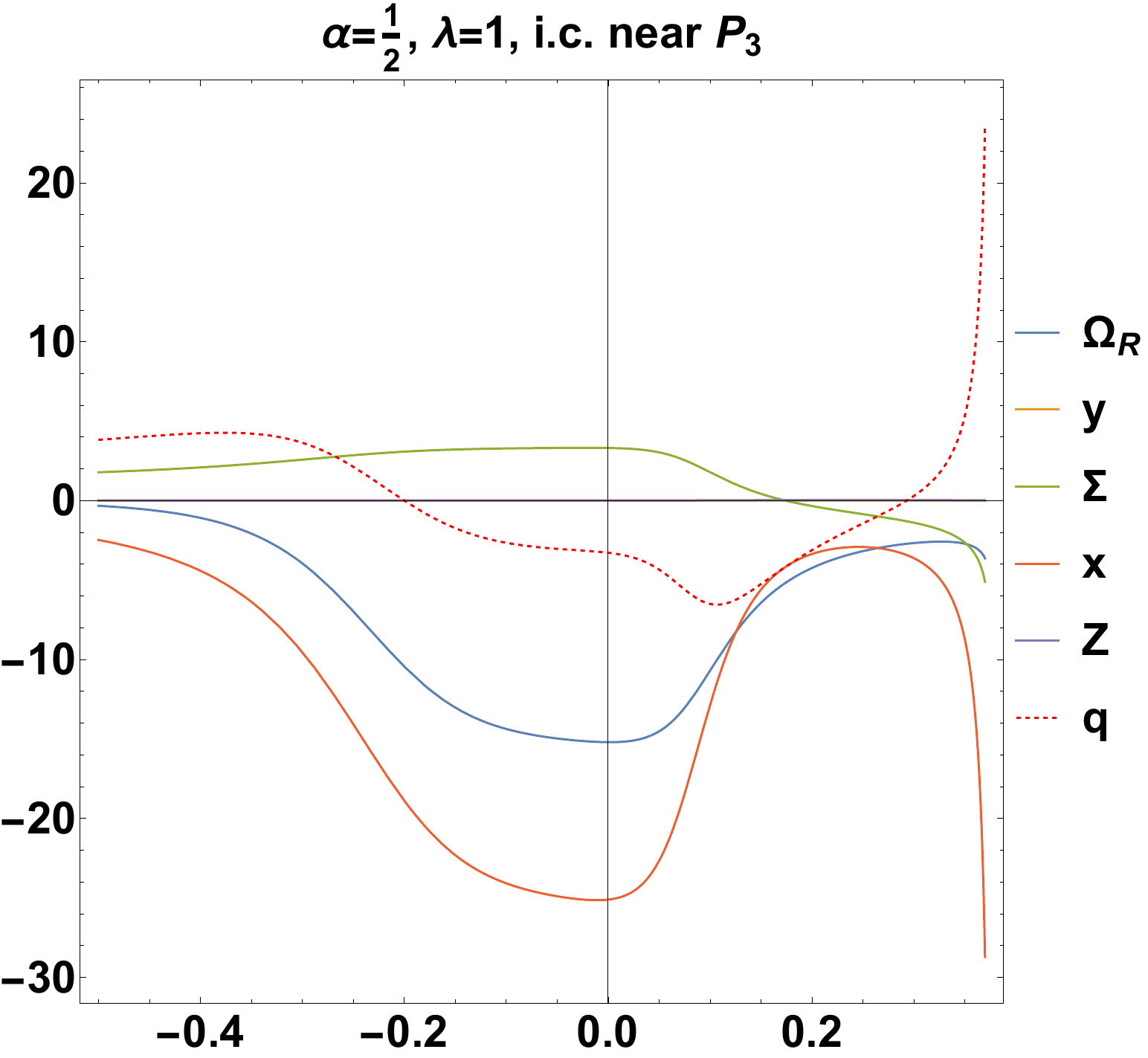}
\includegraphics[width=0.4\textwidth]{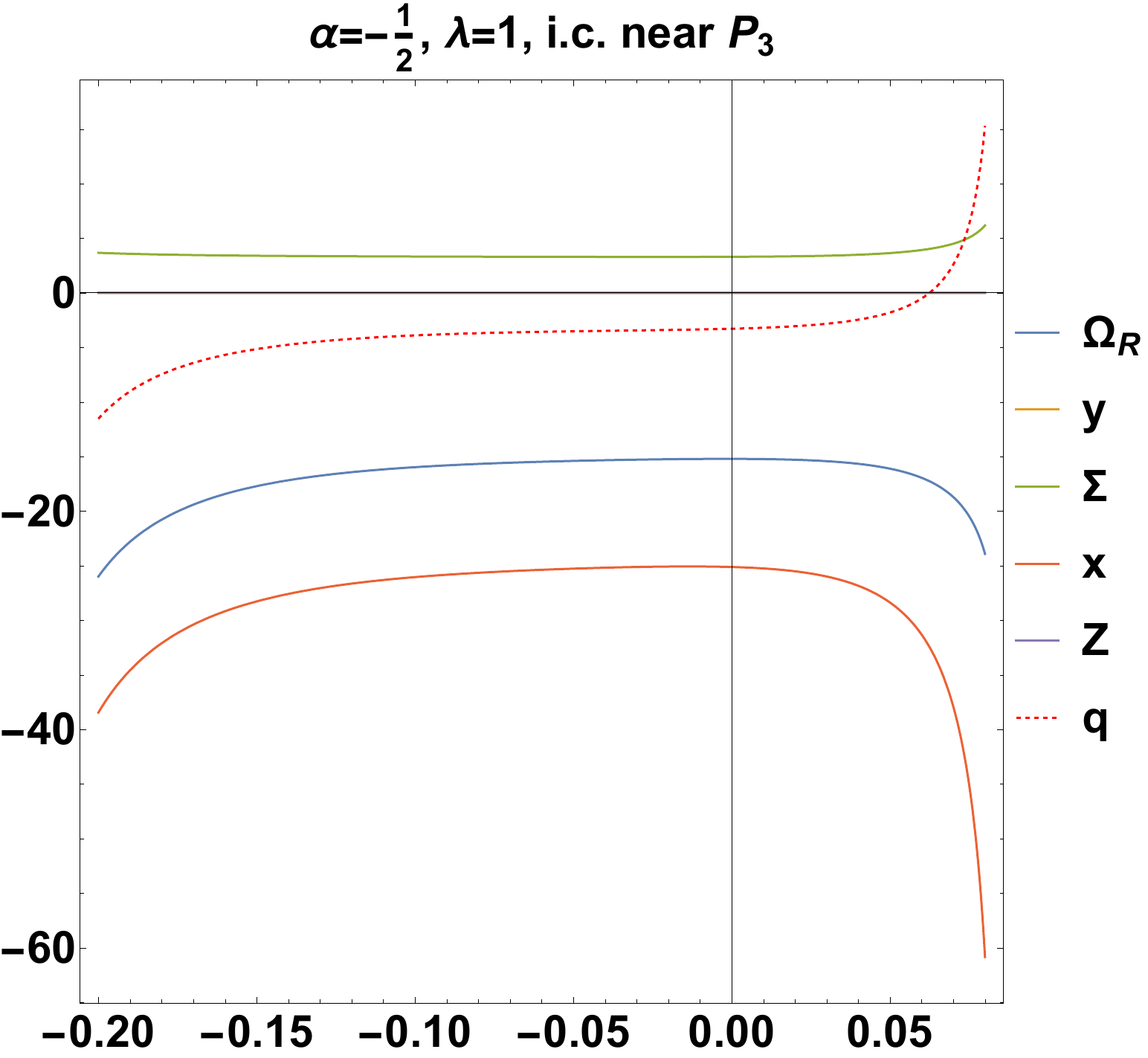}   \caption{Evolution of
$\Omega_{R}, y, \Sigma, x, Z$ evaluated at a numerical solutions of system
\eqref{ds-1-special}-\eqref{ds-4-special} for $\alpha=\pm \frac{1}{2}$ and $\lambda=1$
for initial conditions (i.c.) near the points $P_{2}$ and $P_{3}$ with a
displacement of $\epsilon=\frac{1}{1000}.$ Also present as a red dotted line
is the evolution of the deceleration parameter $q$ evaluated at these points.}%
\label{Fig-2}%
\end{figure}

\subsection{General case}

For the general case with $\alpha$ arbitrary, we introduce the dimensionless
variables
\begin{equation}
\begin{split}
& \Omega_{R}=\frac{ke^{b-2\int H\,dt}}{3H^{2}},\quad y^{2}=\frac{V(\phi)}%
{6H^{2}\phi},\quad\Sigma^{2}=\frac{\dot{b}^{2}}{4H^{2}},\quad x=\frac
{\dot{\phi}}{H\phi},\\
& z=\frac{\dot{\Psi}}{H},\quad w=\frac{\beta\dot{\phi
}e^{-2b-2\int H\,dt}}{6H^{2}\phi\dot{\Psi}}.
\end{split}
\end{equation}

In these variables, the Friedmann equation reads
\begin{equation}
\frac{(k-2\alpha)^{2}xz}{24\alpha}-\frac{x\Omega_{R}}{kz}-\Sigma^{2}%
-w+y^{2}+\Omega_{R}+1=0.\label{friedmann-eq-normalized}%
\end{equation}
This means that we can solve \eqref{friedmann-eq-normalized} for $w$ to obtain
the dynamical system%
\begin{align}
\Omega_{R}^{\prime}= &  \frac{x\Omega_{R}\left(  12\alpha-(k-2\alpha
)^{2}z\right)  }{6\alpha}+2\Omega_{R}\left(  2\Sigma^{2}+\Sigma+y^{2}\right),
\label{ds-5}\\
y^{\prime}= &  \frac{xy\left(  6\alpha(\lambda+1)-(k-2\alpha)^{2}z\right)
}{12\alpha}+y\left(  2\Sigma^{2}+y^{2}+1\right)  \\
\Sigma^{\prime}= &  -\frac{(k-2\alpha)^{2}(\Sigma-1)xz}{12\alpha}%
-\frac{3x\Omega_{R}}{kz}+2(\Sigma-1)^{2}(\Sigma+1)+(\Sigma+2)y^{2}+3\Omega_{R},%
\end{align}%
\begin{equation}
\begin{split}
\left(  -\Sigma^{2}+y^{2}+\Omega_{R}+1\right)  z^{\prime} &  =z\left(
-2(\Sigma-1)^{3}(\Sigma+1)+2\Omega_{R}\left(  (\Sigma-1)^{2}+x\right)
\right)  \\
&  +z\left(  y^{2}((\Sigma-4)\Sigma+(\lambda+1)x+\Omega_{R}+3)+y^{4}\right)
\\
&  -\frac{(k-2\alpha)^{2}}{12\alpha}xz^{2}\left(  -(\Sigma-1)^{2}+y^{2}%
+\Omega_{R}\right)  +\frac{6}{k}\Sigma x\Omega_{R},
\end{split}
\end{equation}%
\begin{equation}
\begin{split}
\frac{12\alpha z}{x} &\left(  -\Sigma^{2}+y^{2}+\Omega_{R}+1\right)
^{2}x^{\prime}  =-(k-2\alpha)^{2}xz^{2}\left(  x\left(  \Sigma^{2}+\lambda
y^{2}+\Omega_{R}-1\right)  \right)  \\
& +(k-2\alpha)^{2}xz^{2}\left(  \left(  -(\Sigma-6)\Sigma+y^{2}+\Omega
_{R}-5\right)  \left(  -\Sigma^{2}+y^{2}+\Omega_{R}+1\right)  \right)  \\
& -\frac{6z}{k}\left(  (k-2\alpha)^{2}\Sigma x^{2}\Omega_{R}\right)
\\
& -6z\left(  4\alpha x\left(  -\Sigma^{2}+y^{2}+\Omega_{R}+1\right)  \left(
\Sigma^{2}+\lambda y^{2}+\Omega_{R}-1\right)  \right)  \\
& -\frac{6z}{k}\left(  2\alpha k\left(  2(\Sigma-1)^{2}+y^{2}\right)  \left(
-\Sigma^{2}+y^{2}+\Omega_{R}+1\right)  ^{2}\right)  \\
& +\frac{(k-2\alpha)^{4}}{6\alpha}(\Sigma-1)x^{2}z^{3}-\frac{72\alpha\Sigma
}{k}x\Omega_{R}\left(  -\Sigma^{2}+y^{2}+\Omega_{R}+1\right),
\end{split}
\end{equation}
where once more the prime means a total derivative with respect the independent variable $\tau=\ln(a)$.

The equilibrium points for the latter system are given by the family
$(\Omega_{R},y,\Sigma,x,z)=(0,0,\Sigma_{c},\frac{24\alpha\left(  \Sigma
_{c}^{2}-1\right)  }{(k-2\alpha)^{2}z_{c}},z_{c})$ where $\Sigma_{c}~$and
$z_{c}\in\mathbb{R}$ are the parameters that define the family. This family is
a normally hyperbolic set of equilibrium points and therefore it has two zero
eigenvalues. We observe that there are not any asymptotic solutions which can
describe anisotropic solutions with nonzero spatial curvature, that is, the
limit of GR is not recovered in this case.

We conclude that the general case is not of physical interest, thus we end the
discussion here.

\section{Concluding remarks}

\label{sec6}

In this study we investigated the asymptotic dynamics for the field equations
in symmetric teleparallel $f\left(  Q\right)  $-theory for Kantowski-Sachs
and Bianchi type III background geometries. The field equations of $f\left(
Q\right)  $-theory are of second-order where the geometrodynamical degrees of
freedom can be attributed to two scalar fields. By using the scalar field
description we were able to write a minisuperspace Lagrangian. From the
minisuperspace approach we observed that, for specific values of some of the free
parameters of theory, some nonlinear terms in the field equations are
eliminated.

To understand the overall evolution of physical parameters in the solution
space, we determined the stationary points of the phase-space and investigated
their stability properties. Employing the Hubble normalization approach, we
transformed the field equations into a system of algebraic-differential
equations. Each stationary point of this system corresponded to an asymptotic
solution, whose stability properties and physical characteristics we
thoroughly examined.

We found that for the general form of the symmetric and teleparallel
connection provided by the theory, the field equations admit asymptotic
solutions describing dynamics similar to that of the Bianchi type I geometry, without recovering the limit of
General Relativity (GR). However, for specific values of the free parameters,
new stationary points emerged, describing the limit of GR and potentially
representing anisotropic and accelerated solutions that could describe the
pre-inflationary epoch of the universe.

From the results of this work it follows that symmetric teleparallel $f\left(
Q\right)  $-theory can describe anisotropic solutions with acceleration.
However, we have considered a power-law function for the $f\left(
Q\right)  $-only and we have not found any future attractor which can describe
an accelerating universe. However, for a more general $f\left(  Q\right)$ function, new
stationary points exist. Finally, we demonstrated how the phase-space analysis
can be utilized to constrain the free parameters of the connection, in order to ensure
the viability of the theory.

\textbf{Data Availability Statements:} Data sharing is not applicable to this
article as no datasets were generated or analyzed during the current study.

\begin{acknowledgments}
A. D. Millano was supported by Agencia Nacional de Investigación y Desarrollo (ANID) Subdirección de Capital Humano/Doctorado Nacional/año 2020 folio 21200837, Gastos operacionales proyecto de tesis/2022 folio 242220121, and VRIDT-UCN. KD acknowledges support by the PNRR-III-C9-2022 Euro call, with project number
760016/27.01.2023. This paper is based upon work from COST Action CA21136
\textit{Addressing observational tensions in cosmology with systematics and
fundamental physics} (CosmoVerse) supported by COST (European Cooperation in
Science and Technology). AG was supported by FONDECYT 1200293. AP thanks the support of VRIDT through
Resolución VRIDT No. 096/2022 and Resolución VRIDT No. 098/2022.
AP thanks ND and the Universidad de La Frontera for the hospitality provided
while part of this work was carried out. AP thanks the support of National Research Foundation of
South Africa.
\end{acknowledgments}

\end{document}